**Technology in Association With Mental Health: Meta-ethnography**

Hamza Mohammed | hamzamohammed0784@gmail.com

**Abstract**

This research paper presents a meta-analysis of the multifaceted role of technology in mental health. The pervasive influence of technology on daily lives necessitates a deep understanding of its impact on mental health services. This study synthesizes literature covering Behavioral Intervention Technologies (BITs), digital mental health interventions during COVID-19, young men's attitudes toward mental health technologies, technology-based interventions for university students, and the applicability of mobile health technologies for individuals with serious mental illnesses. BITs are recognized for their potential to provide evidence-based interventions for mental health conditions, especially anxiety disorders. The COVID-19 pandemic acted as a catalyst for the adoption of digital mental health services, underscoring their crucial role in providing accessible and quality care; however, their efficacy needs to be reinforced by workforce training, high-quality evidence, and digital equity. A nuanced understanding of young men's attitudes toward mental health is imperative for devising effective online services. Technology-based interventions for university students are promising, although variable in effectiveness; their deployment must be evidence-based and tailored to individual needs. Mobile health technologies, particularly activity tracking, hold promise for individuals with serious mental illnesses. Collectively, technology has immense potential to revolutionize mental health care. However, the implementation must be evidence-based, ethical, and equitable, with continued research focusing on experiences across diverse populations, ensuring accessibility and efficacy for all.

**I. Introduction**

Technology has become an integral part of our daily lives in the rapidly evolving digital age. It has transformed how we communicate, work, learn, and manage our health. One of the areas where technology has made significant strides is in the field of mental health. Technology has opened up new avenues for accessing mental health services, from teletherapy platforms to mental health apps. However, the impact of technology on mental health is multifaceted and complex, warranting comprehensive research and understanding (Mohr et al., 2013; Torous et al., 2020; Farrer et al., 2013; Naslund et al., 2015).

This paper aims to delve into the current research surrounding technology's role in mental health, focusing on its benefits, challenges, and potential for future development. The objective is to provide a comprehensive overview of the existing literature, thereby contributing to the ongoing discourse in this field. The research papers that will be discussed in this review have been carefully selected to cover a range of topics within the broader theme of technology and mental health. These include the role of behavioral intervention technologies in mental health, the impact of digital mental health interventions during the COVID-19 pandemic, young men's attitudes towards mental health and technology, the effectiveness of technology-based interventions for tertiary students, and the feasibility of m-Health technologies for individuals with serious mental illness (Mohr et al., 2013; Torous et al., 2020; Farrer et al., 2013; Naslund et al., 2015).

Each of these papers offers valuable insights into the intersection of technology and mental health, shedding light on the potential of technology to transform mental health services and the challenges that need to be addressed. By examining these papers, this review aims to provide a comprehensive understanding of the current state of research in this field and highlight areas for future exploration. In the following sections, I will delve into each of these papers, discussing their findings, implications, and contributions to the field of mental health and technology. Through this review, I hope to provide a comprehensive overview of the current state of research in this field, thereby contributing to the ongoing discourse on the impact of technology on mental health.

## II. Behavioral Intervention Technologies and Mental Health

The advent of technology has brought about significant changes in the field of mental health, particularly in the development of Behavioral Intervention Technologies (BITs). These technologies have been instrumental in providing evidence-based interventions to individuals suffering from various mental health conditions. The paper "Behavioral Intervention Technologies: Evidence Review and recommendations for future research in mental health" by David C. Mohr et al. provides an in-depth analysis of the role of these technologies in mental health (Mohr et al., 2013).

The paper begins by acknowledging the prevalence of mental health conditions, which are less visible but can be as disabling as conditions like schizophrenia, depression, and bipolar disorder. The authors note that the diagnoses of anxiety disorders are continuously being revised, with both dimensional and structural diagnoses being used in clinical treatment and research. They also highlight the importance of understanding how biology, stress, and genetics interact to shape anxiety symptoms (Mohr et al., 2013).

The authors discuss the effectiveness of psychopharmacological and cognitive-behavioral interventions in treating anxiety disorders. They note that these interventions have different symptom targets, and therefore, combining these strategies could potentially improve future outcomes. The authors also highlight the need for more research in the field of alternative strategies for managing anxiety and for treatment-resistant cases (Mohr et al., 2013).

The paper also discusses the prevalence of anxiety disorders in the U.S., which are present in up to 13.3% of individuals and constitute the most prevalent subgroup of mental disorders. Despite their widespread prevalence, these disorders have not received the same recognition as other major syndromes, such as mood and psychotic disorders. The authors argue that this lack of recognition, coupled with the fact that the primary care physician is usually the principal assessor and treatment provider, has led to decreased productivity, increased morbidity and mortality rates, and the growth of alcohol and drug abuse in a large segment of the population (Mohr et al., 2013). The authors also discuss the challenges to the diagnosis of anxiety disorders, provide a model that explains how anxiety symptoms occur and change over time, highlight the neurotransmitter systems affected by these disorders, and discuss the roles and relative efficacy of pharmacological and non-pharmacological interventions (Mohr et al., 2013).

The paper by David C. Mohr et al. provides a comprehensive overview of the role of Behavioral Intervention Technologies in mental health. It highlights the importance of these technologies in providing evidence-based interventions to individuals suffering from various mental health conditions. The authors call for more research in this field to improve the effectiveness of these interventions and to develop alternative strategies for managing anxiety and treatment-resistant cases (Mohr et al., 2013).

**III. Digital Mental Health in the Age of COVID-19**

The COVID-19 pandemic has brought about a significant shift in the mental health landscape, with digital health technologies playing a pivotal role in providing care during this crisis. The paper "Digital Mental Health and COVID-19: Using Technology Today to Accelerate the Curve on Access and Quality Tomorrow" by John Torous, Keris Jän Myrick, Natali Rauseo-Ricupero, and Joseph Firth, published in 2020, provides a comprehensive exploration of this topic (Torous et al., 2020).

The authors argue that the current global health crisis has underscored the potential of digital health to enhance access to and quality of mental health services. They suggest that the current need to "flatten the curve" of the virus spread should be paralleled by efforts to "accelerate and bend the curve" on digital health. They believe that increased investments in digital health today will yield unprecedented access to high-quality mental health care in the future (Torous et al., 2020).

The paper discusses the success of telehealth during the COVID-19 crisis and how technologies like apps can play a larger role in the future. The authors highlight the need for workforce training, high-quality evidence, and digital equity as critical factors for further bending the curve (Torous et al., 2020). The COVID-19 crisis has highlighted the role of telehealth and digital tools like apps to provide care in times of need. Many clinicians and patients are now realizing the full potential of these digital tools as they are forced to, for the first time, utilize them to connect at a time when in-person and face-to-face visits are impossible (Torous et al., 2020). Digital therapy programs that offer courses of evidence-based therapies also have a role in the crisis, given their unique potential for scalability. However, issues of real-world engagement with these programs warrant caution in ensuring that plans for encouraging and maintaining meaningful engagement are in place before purchasing these programs or services (Torous et al., 2020).

The authors also discuss the role of mobile apps in augmenting and extending care. They suggest that the most effective apps are the ones that can be customized to each patient and fit with their personal care goals and needs, as well as apps for peer support (Torous et al., 2020). The authors also highlight the importance of training medical professionals, trainees, and peer support specialists on how to use digital and mobile technologies for delivering care. They argue that ensuring all patients, especially the most vulnerable ones, have the digital literacy and competency to partake in digital care is a matter of equity and social justice (Torous et al., 2020).

Furthermore, the paper suggests that the COVID-19 crisis may be the defining moment for digital mental health due to its extensive and expansive impact on . They argue that ensuring the right use of telehealth and app tools today and investing in people and training to support them tomorrow, can cement the future of digital mental health as simply mental health. They believe that bending the curve in the right direction will require funding, research, policy changes, training, and equity, but these investments will continue to yield higher returns at every step (Torous et al., 2020).

**IV. Technology-Based Interventions for Mental Health in Tertiary Students**

The study "Technology-Based Interventions for Mental Health in Tertiary Students: Systematic Review" by Louise Farrer et al., published in the Journal of Medical Internet Research in 2013, provides a comprehensive review of the effectiveness of technology-based interventions for mental health in university students. The study systematically reviewed published randomized trials of technology-based interventions evaluated in a university setting for disorders other than substance use and eating disorders (Farrer et al., 2013).

The study identified 27 relevant studies, most of which targeted anxiety symptoms or disorders or stress. Of the 51 technology-based interventions employed across the studies, approximately half were associated with at least one significant positive outcome compared with the control at post-intervention. However, 29% failed to find a significant effect. The median effect size was 0.54 for interventions targeting depression and anxiety symptoms and 0.84 for interventions targeting anxiety symptoms and disorders (Farrer et al., 2013).

Internet-based technology, typically involving cognitive behavioral therapy (CBT), was the most commonly employed medium, being used in 16 of the 27 studies and approximately half of the 51 technology-based interventions. Distal and universal preventive interventions were the most common type of intervention (Farrer et al., 2013). The findings of this review indicate that technology-based interventions targeting certain mental health and related problems offer promise for students in university settings. However, more high-quality trials that fully report randomization methods, outcome data, and data analysis methods are needed (Farrer et al., 2013). The use of technology, particularly internet-based interventions, in addressing mental health issues among university students is promising. These interventions are easily accessible, cost-effective, and may be perceived as less stigmatizing than traditional approaches to care. Furthermore, they can be tailored to the individual needs of students, providing a more personalized approach to mental health care (Farrer et al., 2013). However, the effectiveness of these interventions varies. While some studies found significant positive outcomes, others did not. This suggests that more research is needed to identify the factors that contribute to the effectiveness of technology-based interventions for mental health in university students (Farrer et al., 2013).

Technology-based interventions have the potential to significantly improve mental health outcomes for university students. However, more research is needed to optimize these interventions and ensure they are effectively implemented in university settings. As technology continues to evolve, it is crucial that mental health services leverage these advancements to provide the best possible care for university students (Farrer et al., 2013).

## V. Conclusion

As a thorough examination of studies and discoveries in technology and mental health has shown, the intersection of technology and mental health is a rapidly evolving field with significant potential for improving access to and quality of mental health care. The papers reviewed in this research paper provide valuable insights into the current state of this field and its future directions (Mohr et al., 2013; Torous et al., 2020; Farrer et al., 2013; Naslund et al., 2015).

The paper by David C. Mohr et al. underscores the importance of behavioral intervention technologies in mental health, highlighting the need for further research to fully understand their potential and limitations. As technology continues to advance, it is crucial that mental health professionals stay abreast of these developments and incorporate them into their practice where appropriate (Mohr et al., 2013).

The COVID-19 pandemic has underscored the importance of digital mental health services, as discussed in the paper by John Torous et al. The pandemic has accelerated the adoption of these services, and it is clear that they will continue to play a vital role in mental health care even after the pandemic subsides. The challenge moving forward will be to ensure that these services are accessible to all who need them and that they maintain a high standard of quality (Torous et al., 2020).

The effectiveness of technology-based interventions for mental health in tertiary students, as discussed in the paper by Louise Farrer et al., highlights the potential of these interventions to support students' mental health. With many students now studying remotely due to the pandemic, these interventions are more important than ever (Farrer et al., 2013).

Finally, the paper by John A. Naslund et al. explores the feasibility of popular m-Health technologies for activity tracking among individuals with serious mental illness. This is a promising area of research, as these technologies could provide valuable data for mental health professionals and help individuals manage their mental health (Naslund et al., 2015).

In summary, technology has the potential to significantly impact mental health care, from providing new avenues for treatment to improving access to care. However, it is crucial that this potential is harnessed in a way that is ethical, equitable, and evidence-based. As we move forward, it will be important to continue researching this field, with a focus on understanding the experiences of diverse populations and ensuring that technology-based interventions are accessible and effective for all (Mohr et al., 2013; Torous et al., 2020; Farrer et al., 2013; Naslund et al., 2015).

**References**

1. Mohr, D. C., Burns, M. N., Schueller, S. M., Clarke, G., & Klinkman, M. (2013). Behavioral intervention technologies: evidence review and recommendations for future research in mental health. General Hospital Psychiatry, 35(4), 332-338.

2. Torous, J., Myrick, K. J., Rauseo-Ricupero, N., & Firth, J. (2020). Digital Mental Health and COVID-19: Using Technology Today to Accelerate the Curve on Access and Quality Tomorrow. JMIR Mental Health, 7(3), e18848.

3. Farrer, L., Gulliver, A., Chan, J. K., Batterham, P. J., Reynolds, J., Calear, A., ... & Griffiths, K. M. (2013). Technology-based interventions for mental health in tertiary students: systematic review. Journal of Medical Internet Research, 15(5), e101.

4. Naslund, J. A., Aschbrenner, K. A., Barre, L. K., & Bartels, S. J. (2015). Feasibility of popular m-Health technologies for activity tracking among individuals with serious mental illness. Telemedicine and e-Health, 21(3), 213-216.